\documentclass[a4paper]{article}
\usepackage{flushend}
\usepackage{makecell}

\usepackage{INTERSPEECH2021}

\title{ProsoBeast Prosody Annotation Tool}
\name{Branislav Gerazov$^1$ and Michael Wagner$^2$}
\address{
  $^1$Faculty of Electrical Engineering and Information Technologies\\
  Ss Cyril and Methodius University in Skopje, Macedonia\\
  $^2$McGill University, Canada }
\email{gerazov@feit.ukim.edu.mk, chael@mcgill.ca}

\begin{document}

\maketitle
\begin{abstract}
The labelling of speech corpora is a laborious and time-consuming process.
The ProsoBeast Annotation Tool seeks to ease and accelerate this process by providing an interactive 2D representation of the prosodic landscape of the data, in which contours are distributed based on their similarity.
This interactive map allows the user to inspect and label the utterances.
The tool integrates several state-of-the-art methods for dimensionality reduction and feature embedding, including variational autoencoders.
The user can use these to find a good representation for their data.
In addition, as most of these methods are stochastic, each can be used to generate an unlimited number of different prosodic maps.
The web app then allows the user to seamlessly switch between these alternative representations in the annotation process.
Experiments with a sample prosodically rich dataset have shown that the tool manages to find good representations of varied data and is helpful both for annotation and label correction.
The tool is released as free software for use by the community.
\end{abstract}
\noindent\textbf{Index Terms}: prosody annotation, labelling, dimensionality reduction, latent space, variational autoencoders

\section{Introduction}

Speech corpora are an essential part of speech research.
In the analysis of prosody, annotation of the recorded material is a necessary part of the process of corpus creation. 
The purpose of annotation is to assign labels from a predetermined set that best characterise a particular aspect of the utterance.
In prosody research this could be the type of intonation contour present in the utterance, but could also involve other variables, such as the prosodic correlates of syntactic junctures, or the  presence of emphasis in certain parts of an utterance. 
Tune, grouping and emphasis are three in principle orthogonal linguistic dimensions of prosodic organization, and there may be additional paralinguistic factors in addition \cite{ladd2008intonational}.
The annotation process of data on these prosodic phenomena often necessitates manual assignment of the labels by at least one expert, one utterance at a time.
This is a tedious, time-consuming and expensive process.

We introduce the ProsoBeast Annotation Tool --- a web app that seeks to facilitate the annotation process, making it easier to annotate similar intonation contours by clustering them in a 2D representation.
The representation is fully interactive and it allows the annotator to closely inspect and compare the contour shapes and listen to the utterances when making a decision on the label.
Projecting high-dimensional data to a 2D space is a lossy process that is not guaranteed to provide a useful representation.
To this end, the tool integrates different state-of-the-art methods for dimensionality reduction including: PCA, t-SNE, Variational Autoencoders, and Recurrent Variational Autoencoders.
The tool is made available as free software.\footnote{\tiny{\url{https://github.com/prosodylab/prosobeast-annotation-tool}}}

The purpose of the tool is manyfold. 
It can speed up the process of assigning labels to prosodic contours if they are already `ordered' by similarity w.r.t. relevant acoustic dimensions.
It could help reduce certain types of errors, for example misclassifying a tune because one has just listened to a radically different intonation tune that differed along many dimensions. 
It is also useful in finding potential errors in an existing annotation, since tunes that are annotated with different labels than their immediate neighbours, may have been misclassified. 
Finally, the tool can be useful in assessing whether the acoustic features currently used capture the perceptually relevant information. 
If the automatically generated map of the intonation contours does not align with the annotations, this could reveal that relevant acoustic cues have not been measured, or that the acoustic dimensions are somehow not represented in a way that reflect the way our perceptual system interprets them. 
We expect therefore that using the tool will also help us detect shortcomings in our understanding of the intonational phenomena involved. 

\section{Dimensionality Reduction Methods}

The selection of dimensionality reduction methods that are integrated within the Prosody Annotation Tool, allows the users to obtain different 2D projections of their data.
Moreover, since most of these methods contain stochastic elements, the user can process the data multiple times with each to obtain different representations.
Three of the proposed methods work with fixed length data, and one works with variable length input.
All of the methods are accessible through the web app user interface.

\textbf{PCA.}
Principal component analysis is a deterministic algorithm that calculates a projection of the data onto an orthogonal coordinate system that optimally describes its variance \cite{pearson1901liii}.
The axes of the new orthonormal basis are the principle components of the data.
The first of these is the axes that maximizes the variance of the projected data, and each successive one is the orthogonal axis to all the previous axes that maximizes the variance in the projected data.
Although PCA is not a dimensionality reduction technique \emph{per se}, it is often used to this end by projecting the data only onto the first few components.

\textbf{t-SNE.}
t-distributed stochastic neighbour embedding is a nonlinear dimensionality reduction method for visualisation that maps similar high-dimensional objects onto nearby points in a 2D projection space, whilst mapping dissimilar objects to distant points in this space with high probability \cite{hinton2002stochastic, van2008visualizing}.
The t-SNE is able to visually cluster data based on its local structure, and to disentangle complex data with several manifolds.
The algorithm is stochastic, i.e. it will output a different projection of the data for the same choice of hyperparameters if different random seeds are used.
We use this to our advantage in the annotation tool, as users can rerun the algorithm to get new visualisations of the data.

\begin{figure*}[th]
  \setlength\belowcaptionskip{-10pt}
 \centering
\centering
 \hfill
 \includegraphics[height=18em]{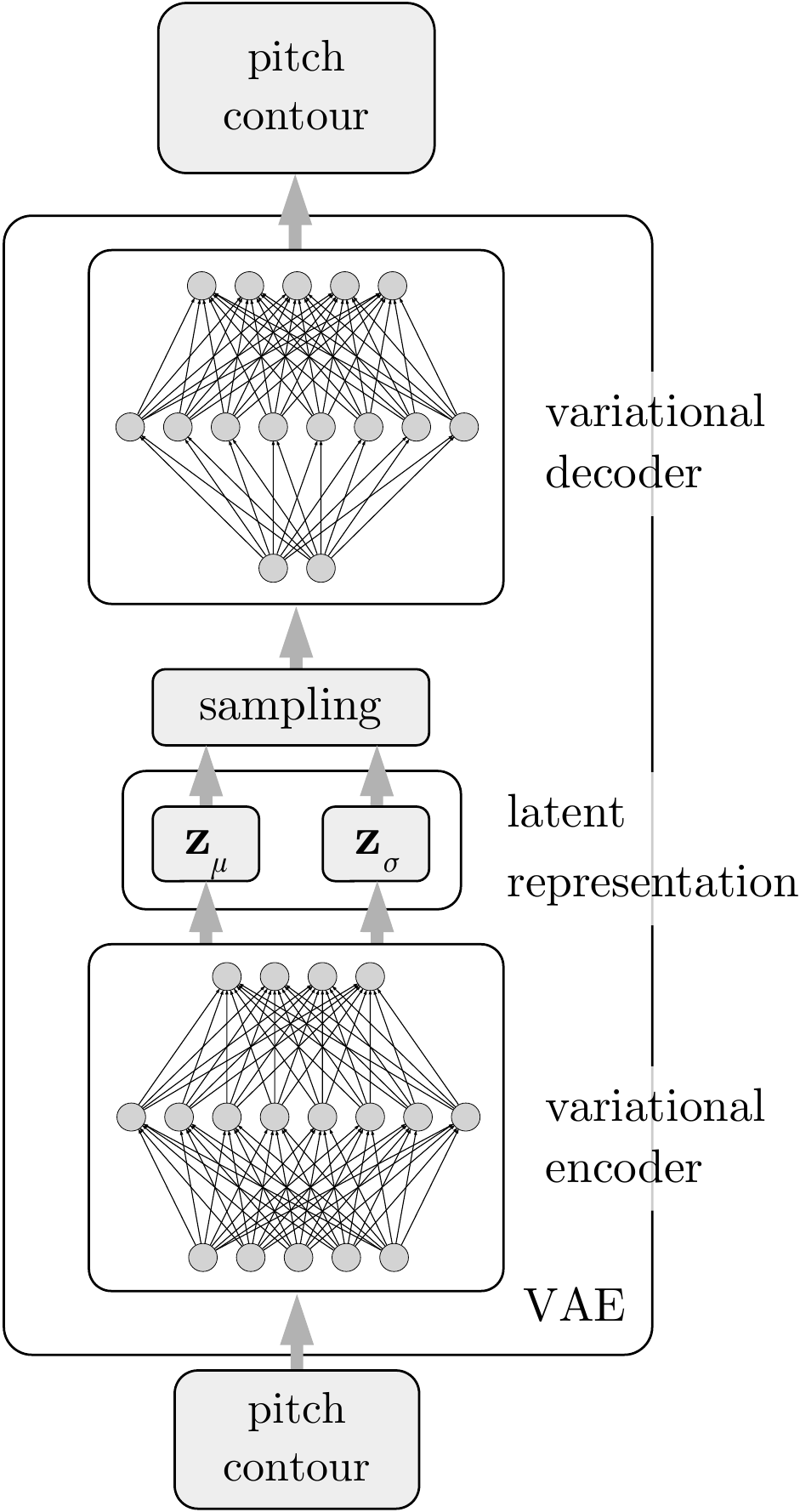}
 \hfill
 \includegraphics[height=18em]{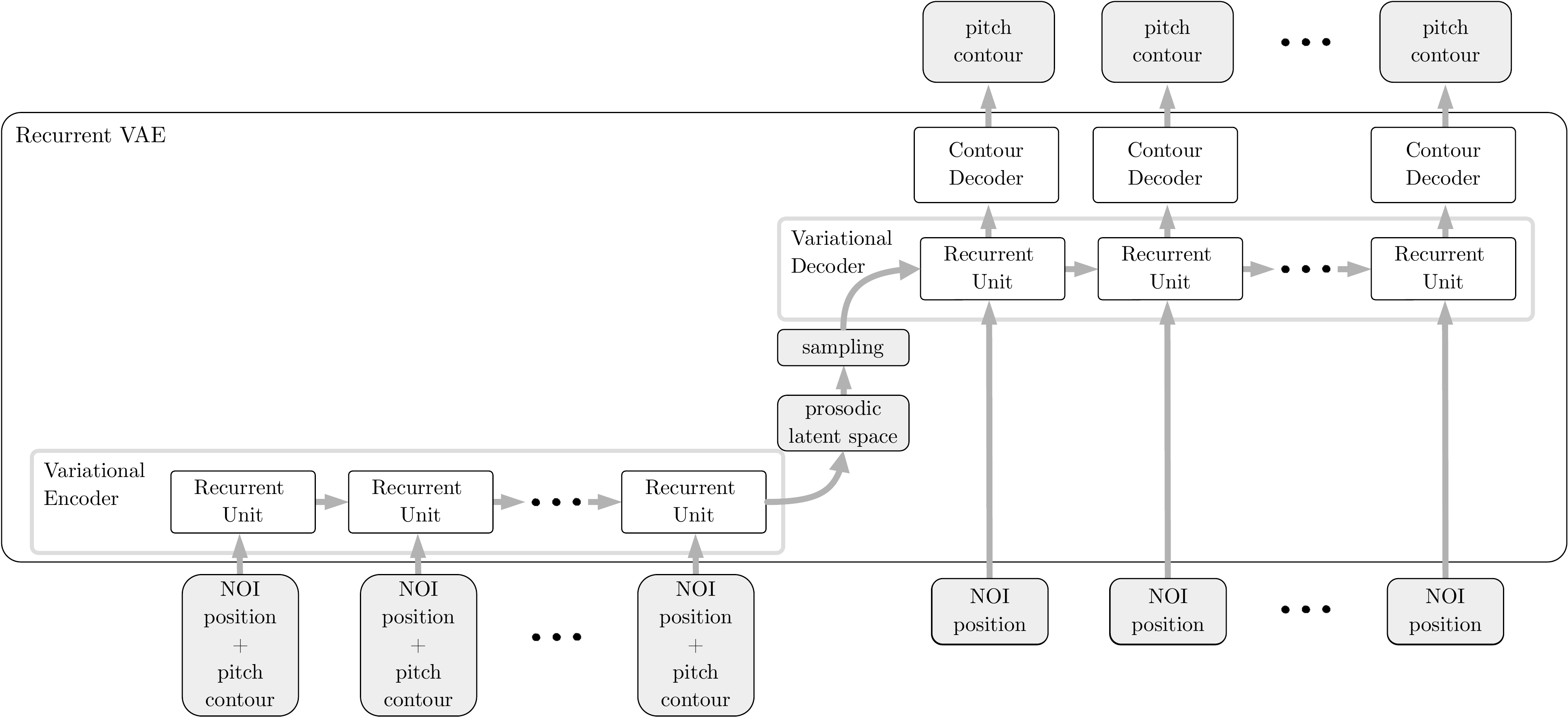}
 \hfill\null
 \caption{Architecture of the Variational autoencoder (VAE) and the recurrent variational autoencoder (RVAE).}
  \label{fig:vae}
\end{figure*}

\textbf{VAE.}
Variational autoencoders are an extension of the classical autoencoders --- deep neural networks built to learn a compressed representation of the input data, i.e. an encoding scheme \cite{cottrell1988principal, deng2010binary}, as well as for
learning meaningful signal representations \cite{socher2011semi, liou2014autoencoder, ap2014autoencoder, obin2018sparse}.
VAEs differ in that they learn a probabilistic projection of the data onto a representation space, also called a latent space, which is continuous and allows for sampling and interpolation \cite{kingma2013auto}.
The VAE architecture shown in Fig.~\ref{fig:vae}, consists of an \emph{encoder} network responsible for projecting the input onto the learned latent space, and a \emph{decoder} network which does the opposite and transforms back the original input. 
In fact, in VAEs the encoder maps the input to a probability distribution of the latent representation conditioned on the input.
The decoder on the other hand, is used to train the encoder to preserve the information in the input relevant to its shape.
In VAEs the encoder and decoder are built using a multilayered feedforward deep neural network (DNN).

\begin{figure}[t]
    \centering
    \vspace{5pt}
    \includegraphics[width=0.6\linewidth]{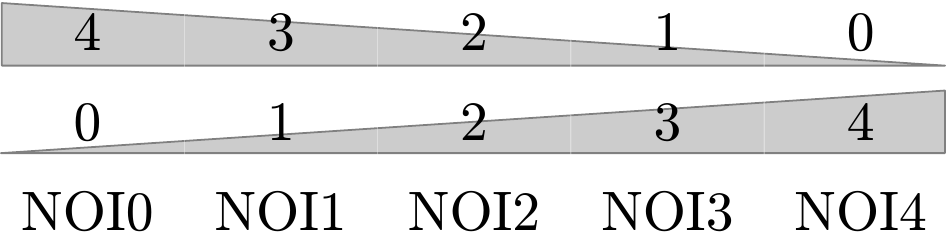}
    \caption{Example position ramps for a pitch contour comprising 5 time steps. The first ramp encodes the distance to the end of the contour, and the second encodes the distance from its beginning.}
    \label{fig:ramps}
    \vspace{-10pt}
\end{figure}

\textbf{RVAE.}
The limitation of the VAE architecture is that it can only be used with fixed length data.
Recurrent VAEs (RVAEs) extend the VAE concept to sequence-to-sequence models that are inherently designed to handle variable length data~\cite{fabius2014variational}.
For the purposes of the annotation tool, we designed and implemented an RVAE that uses positional encodings to augment the encoder input and to drive the decoder, shown in Fig.~\ref{fig:vae}.
In this way, we circumvent the need of having the conclusion of the decoding process to depend on a prediction of an end-of-sequence symbol.
The encodings comprise the two position ramps shown in Fig.~\ref{fig:ramps}.
This is a generalisation of the Variational Recurrent Contour Generator introduced in the Variational Prosody Model \cite{gerazov2018variational}.

In the case of VAEs and RVAEs, more than 2 dimensions might be needed to capture the complexity of the data in the latent space representation.
To facilitate the visualisation of higher-dimensional latent spaces, the projected data can additionally be processed with the t-SNE to bring it down to 2D.

\section{User Interaction}

The ProsoBeast Annotation Tool is built as a lightweight WSGI web application using the Python Flask framework. 
Users can run the application on their local machine by running a Flask server or host it on a web server.

\begin{figure}[t]
    \centering
    \includegraphics[width=\linewidth]{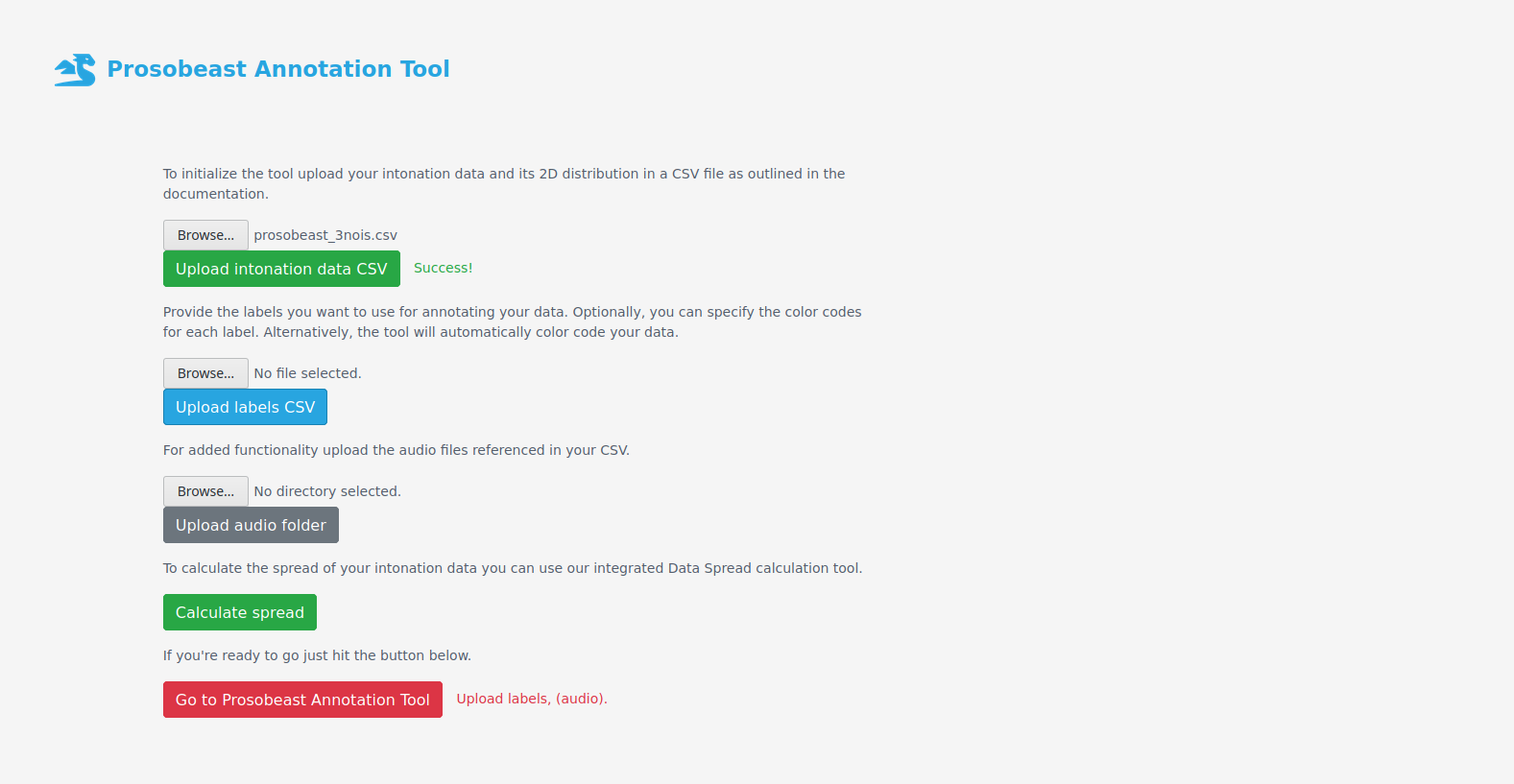}
    \caption{ProsoBeast annotation tool Initialisation screen.}
    \label{fig:init}    
    \vspace{-10pt}
\end{figure}
\textbf{Initialisation screen.}
The init screen shown in Fig.~\ref{fig:init} allows the user to upload their data using a CSV file. 
The CSV file should contain the filename, information about the utterance, the given label, the location in a 2D representation (optional), and the $f_0$ values.
Contours for which labels are not provided are automatically assigned a `No label'.

Another CSV file is needed that specifies the full set of labels that will be used for annotation.
This file can optionally specify the colour codes to use for each label. 
If colour codes are not provided, the tool will automatically generate ones, using grey for the `No label' class.
Finally, the init screen allows users to optionally upload an audio folder containing the utterance files referenced in the data CSV. 
This will allow playback of the audio files in the interactive visualisation.

\begin{figure}[t]
    \centering
    \includegraphics[width=\linewidth]{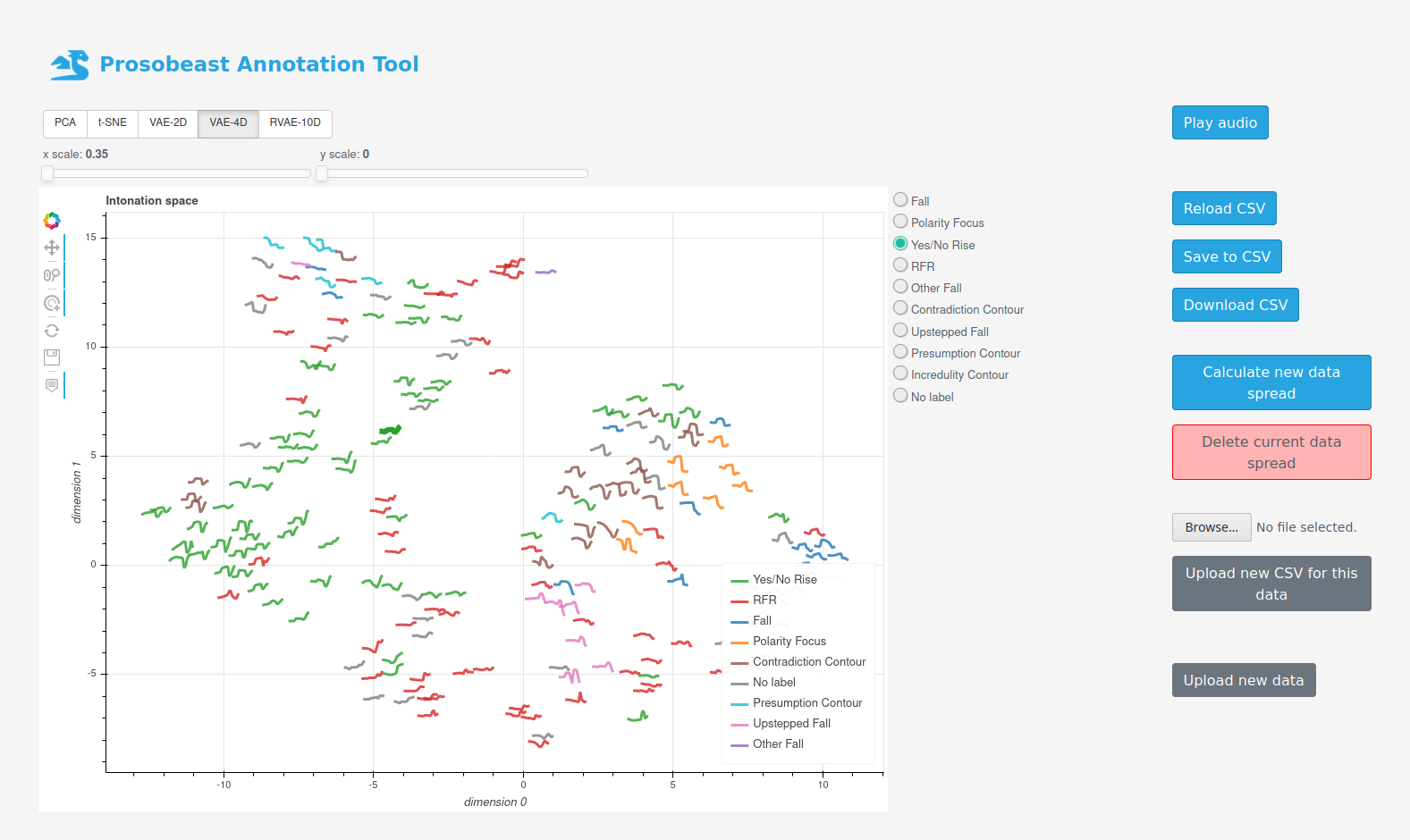}
    \caption{
    ProsoBeast annotation tool interface.}
    \label{fig:app}
    \vspace{-10pt}
\end{figure}

\textbf{Data spread calculation screen.}
The methods for calculating the prosody space projections data can be accessed through the Data spread screen.
The user is automatically redirected to the data spread screen if no location data is provided in the data CSV. 
This screen can also be accessed via the main screen to calculate new data spreads for the data during annotation.

If the prosody data is of fixed length, then the user can use the PCA, t-SNE, and VAE methods.
There are two VAE projection settings --- one for a 2D latent space and one for a 4D latent space.
For variable length data the user can use the RVAE with a preset 10D latent space.
For both the 4D VAE and 10D RVAE, the latent space projection is processed via t-SNE to obtain the final 2D visualisation.
To streamline the use of the tool, the default hyperparameters can be adjusted only via code.
The PCA and t-SNE calculations are implemented using scikit-learn~\cite{scikit-learn}.
The VAE and RVAE models have been implemented using PyTorch for the VAE model training.

\textbf{Main screen.}
The main mode of user interaction in the main screen shown in Fig.~\ref{fig:app} is the interactive plot of the 2D representation of the data. 
It shows each utterance in the dataset represented by its pitch contour, coloured according to its label, if one has been assigned, or grey otherwise.
The plot can be zoomed and scrolled to examine densely packed clusters more closely.
When hovering over the contours, the filename of the utterance is given as well as the adjoining information. 
The contours can be clicked to hear the audio recording.
If multiple representations are calculated, the user can switch between them via the buttons above the plot.
Also, the scaling of the contours for each representation can be adjusted via the two sliders.
The plot is rendered with JavaScript using the Bokeh.

Once a contour is selected it can be assigned a label using the radio buttons on the right of the plot.
The contour remains selected across the generated representations, so the user can switch between them to find the one with appropriate clustering for that contour.
On the far right, the tool has buttons that the user can use to replay the contour's audio, restart labelling by reloading the saved label CSV, save the current labels in a label CSV, download the saved label CSV, calculate a new data spread, delete the current data spread, upload new labels for the current data, or go back to the init screen and upload new data.

\section{Experiments}

To asses the usefulness of the annotation tool we evaluated it using a sample prosody-rich dataset comprising utterances of variable lengths.
We also used the dataset to evaluate the different algorithms for calculating the 2D representations.

\subsection{Dataset}

As a sample dataset we used a subset of the Intonation Bestiary dataset~\cite{wagnergoodhue21}.
The dataset consists of productions of fixed carrier utterances that speakers were asked to pronounce with different intonation patterns in order to convey a certain meaning within a dialogue setting~\cite{goodhue2016toward}.
The speakers responded to a prerecorded utterance, and were given `stage directions' about the precise intent of the utterance they were recording.
There were three kinds of intent: 
\emph{i}) utterance that contradicted a claim the interlocutor had made, 
\emph{ii}) utterances that insinuated an implication in addition to what is asserted, and 
\emph{iii}) utterances that expressed incredulity about what is asserted. 
These were intended to elicit different intonations: for contradiction --- a falling contour with polarity focus~\cite{goodh18} and the contradiction contour~\cite{liber74}, for insinuations --- the rise-fall-rise contour~\cite{ward85},  and for incredulous utterances --- question contours and also the incredulity contour~\cite{hirsc92}. 
Speakers produced a wide variety of contours including the expected ones, except the incredulity contour, which was exceedingly rare if present at all. 
Instead, speakers almost always used some version of a polarity-question rise to convey incredulity.
\cite{goodh18}  categorized the data into different tune categories based on several annotations, which were consolidated based on consensus taking into account
divergencies and consistencies.
Besides being prosodically rich, the dataset is also interesting because it contains data with varying length, i.e. the contours are superimposed on carrier sentences with 3 and 4 content words in a 2:1 ratio. 
Although this is only limited variability, it still allowed us to test run the tool's 2D representation capabilities for variable length data.

We extracted the pitch contours from the data using Kaldi's pitch extractor \cite{ghahremani2014pitch}, in a two pass process as suggested by Hirst~\cite{hirst2011analysis}.
In the first pass we use Praat's~\cite{boersma2001praat} default pitch bounds of 75 and 600~Hz.
From all the extracted pitch values we only keep the ones whose Probabilities of Voicing (POV) pass a loose threshold of 0.2 out of 1. 
We then determine the 1st and 3rd quartiles of the kept pitch value distribution for each speaker. 
These are scaled with 0.25 and 1.5 to get the $f_0$ bounds with which we perform the second pass.
We found that this approach led to satisfactory results in pitch extraction for most of the speakers in the dataset, dealing efficiently with octave jump errors.
The $f_0$ contours were finally normalised to the log scale:
\begin{equation}
f_{0 \, \text{norm}} = 240 \log_2 \frac{f_0}{f_{0 \, \text{ref}}}
\end{equation}
where $f_{0 \, \text{ref}}$ is 
the reference pitch based on
the maximum of the Kernel Density Estimation function for each speaker 
\cite{rosenblatt1956remarks}.

We sampled each pitch contour at 10, 30, 50, 70 and 90\% of the duration of the stressed syllable nuclei of the carrier content words, which we call \emph{nuclei of interest} (NOI).
This approach, even though disregarding chunks of the utterance melody, allowed us to: 
\emph{i}) keep the its most perceptually relevant parts, and 
\emph{ii}) to represent the data with a fixed number of points, i.e. 15 and 20 samples for the 3 and 4 content word sentences, at 5 samples per NOI.
Based on the extracted pitch data, we decided to keep only the utterances that had at least 50\% of their NOI pitch values with POV greater than 0.2.
The selection process resulted with 205 utterances with 3 NOI and 96 utterances with 4 NOI, or a total of 301 utterances, as shown in Table~\ref{tab:data}.
For reproducibility, the data is made available for open use.\footnote{\url{https://zenodo.org/record/4660054}}

\begin{table}[h]
    \caption{Spread of the sample dataset.}
    \label{tab:data}
    \centering
    \begin{tabular}{ c c c c }
        \toprule
        \makecell{NOI per \\ utterance} & Utterances & Total NOI & \makecell{Total pitch \\ samples} \\   
        \midrule
        3 & 204 & 612 & 3060 \\
        4 & 96 & 384 & 1920 \\
        \midrule
        Total & 300 & 996 & 4980 \\   
        \bottomrule
    \end{tabular}
    \vspace{-10pt}
\end{table}

\subsection{2D Projections}

We applied the PCA, t-SNE and VAE dimensionality reduction techniques to the 3 NOI subset of the sample dataset, and the RVAE to the whole dataset.
For the t-SNE we used a perplexity of 30, random initialisation and 5000 iterations.
The 2D and 4D latent space VAE we used 2 hidden layers of 128 neurons each in the encoder and decoder.
We use Gated Recurrent Units (GRUs) for the RVAE model with a hidden layer size of 512.
For all neural models we use a 0.001 learning rate, $10^{-4}$ L2 regularisation, and the Adam optimiser~\cite{kingma2014adam}.
We don't fix the random seed so that the user can repeatedly use the methods to generate new prosody latent spaces.
We made this choice of hyperparameters through experimentation with the sample dataset.
The user can optimise these to their data using the data spread master script included in the code.

\begin{figure}[th!]
    \centering
    \includegraphics[width=.66\linewidth]{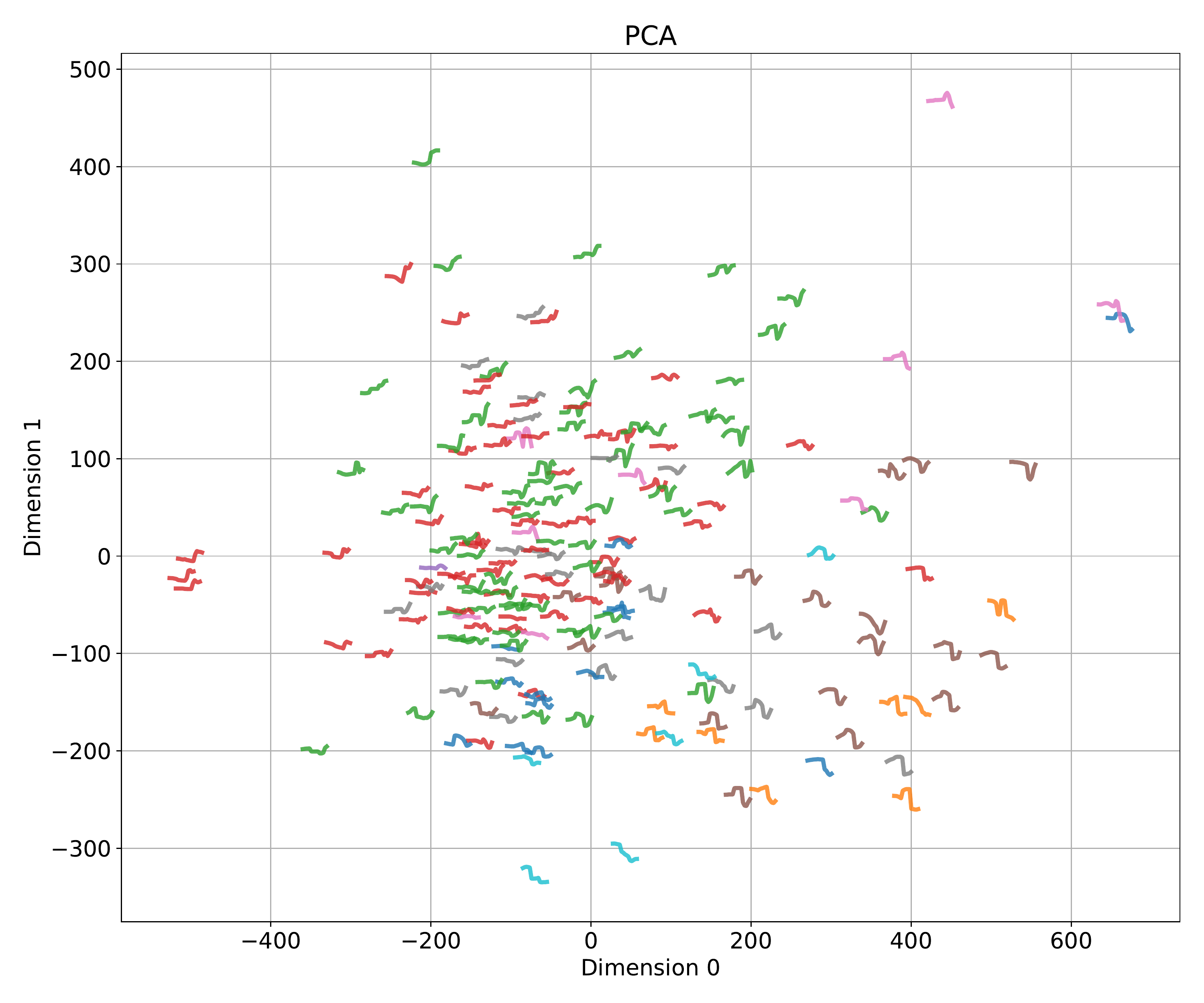}
    
    \includegraphics[width=.66\linewidth]{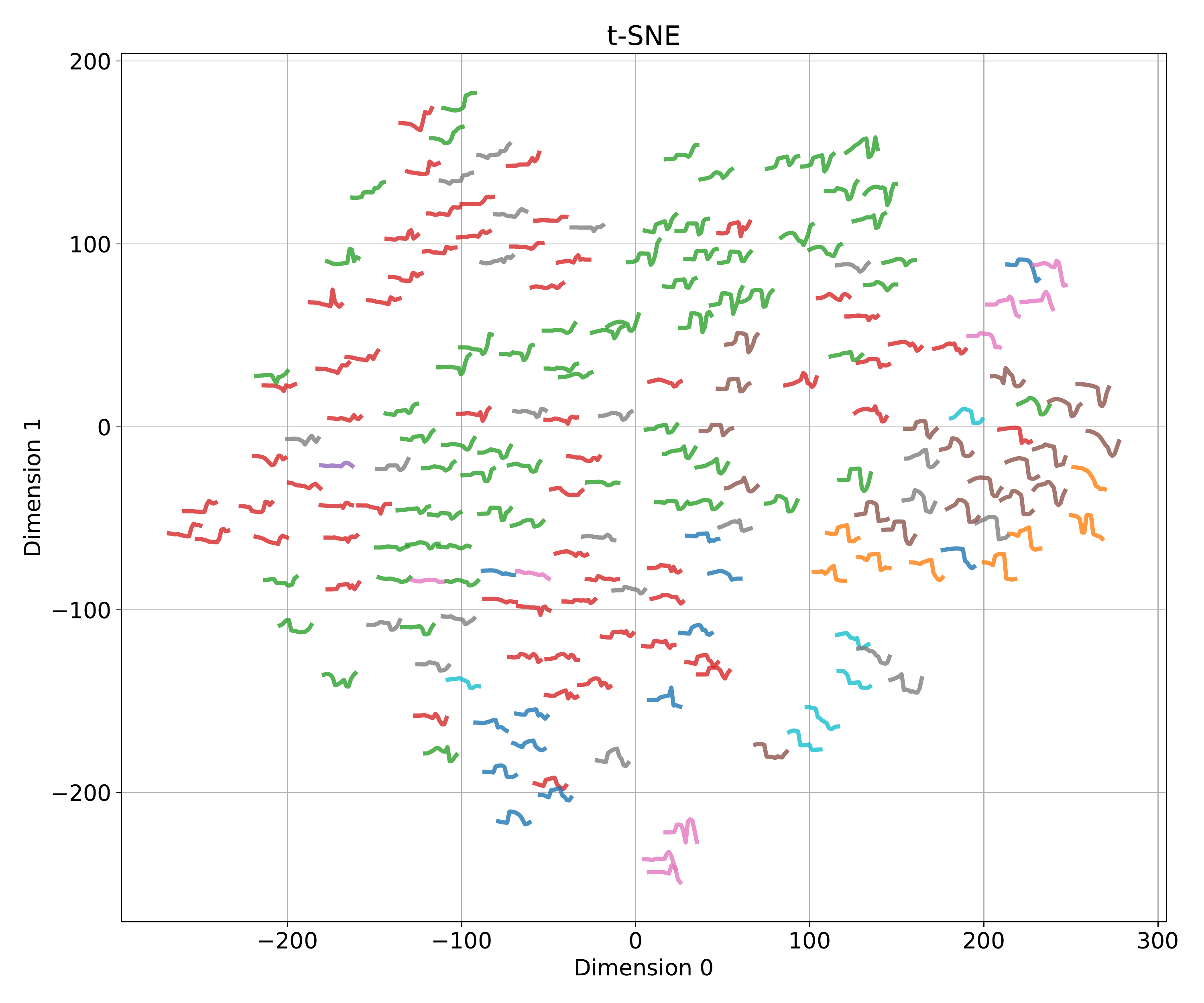}
    
    \includegraphics[width=.66\linewidth]{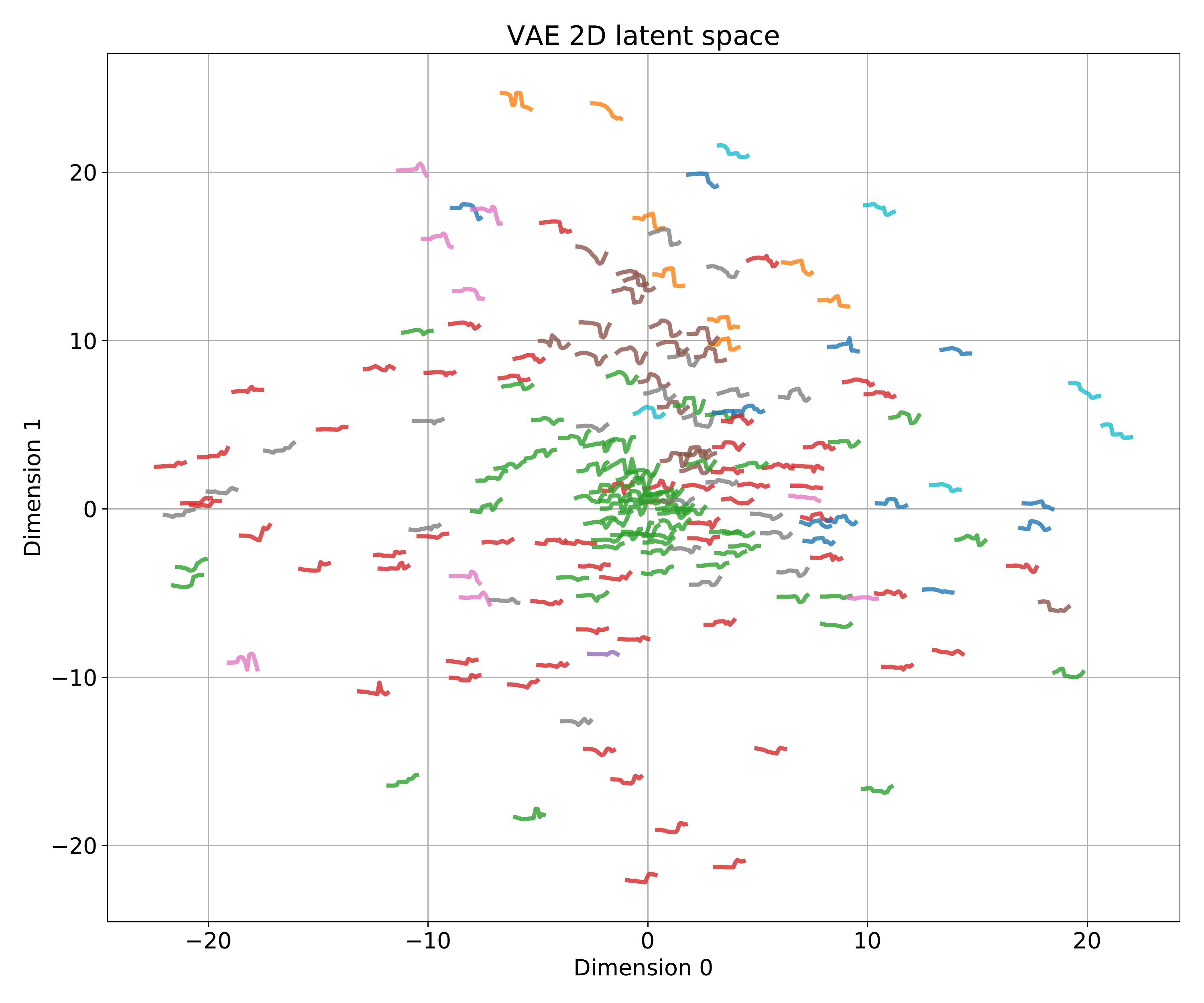}
    
    \includegraphics[width=.66\linewidth]{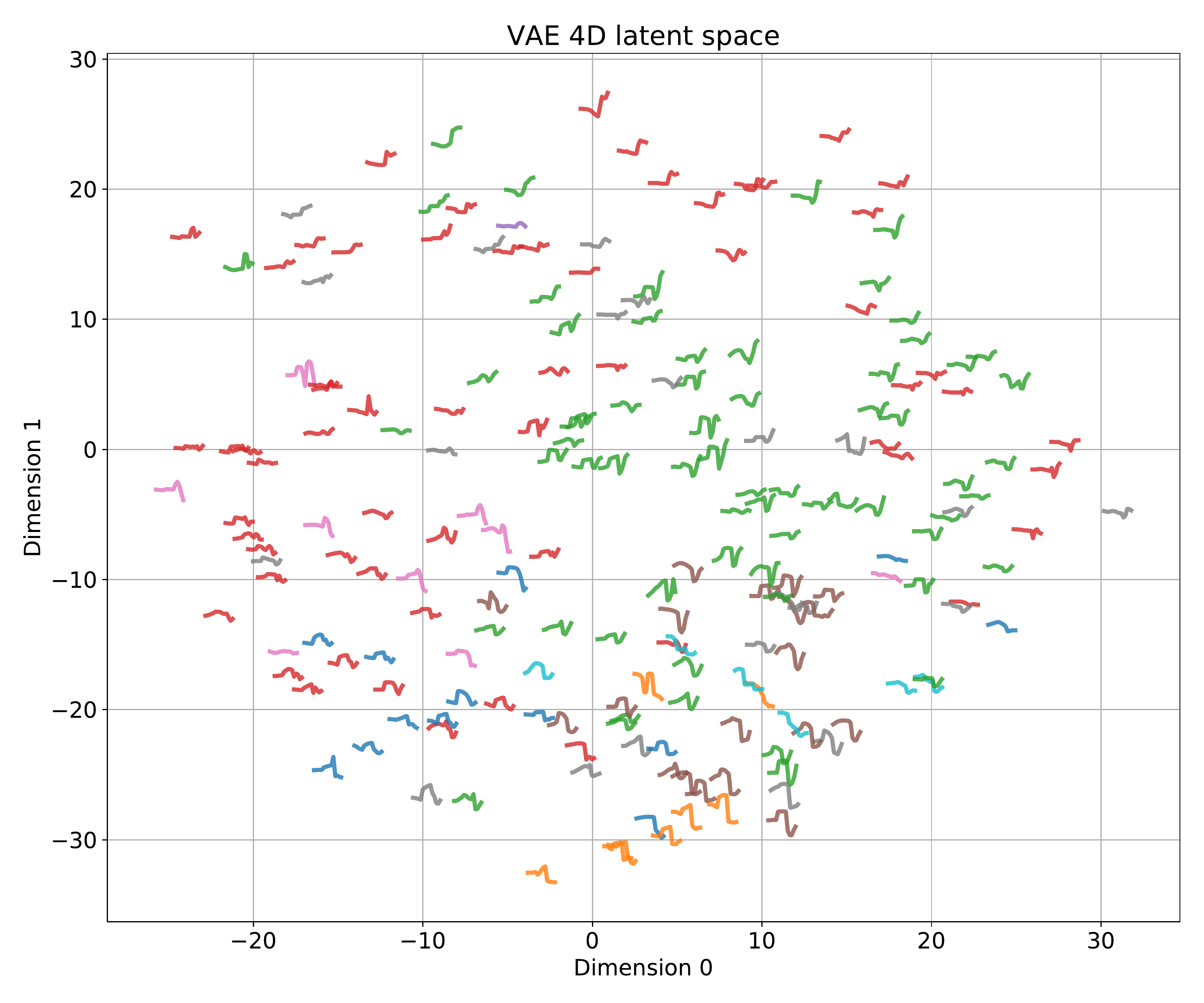}
    
    \includegraphics[width=.66\linewidth]{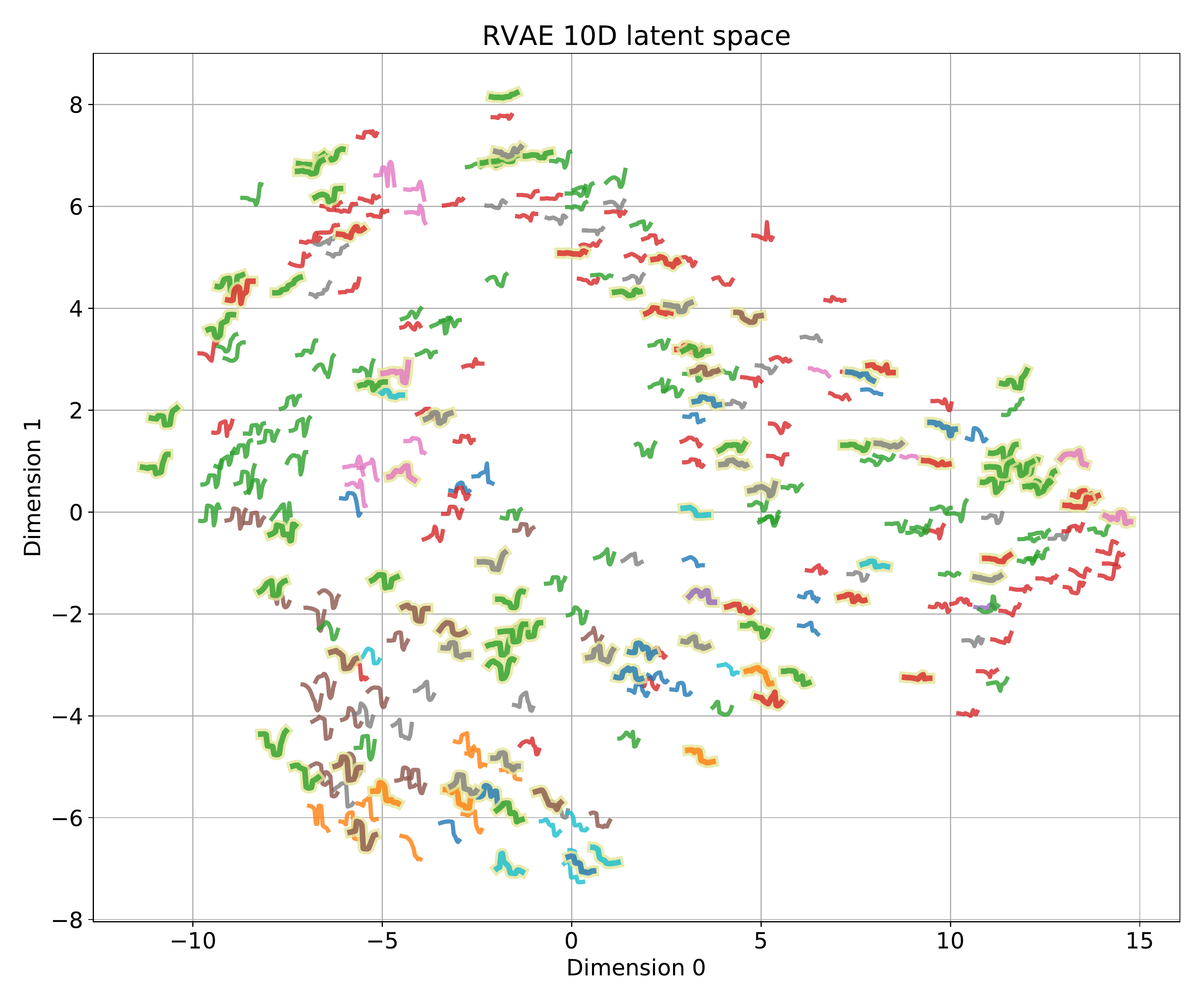}
    \caption{Data spreads calculated with the PCA, t-SNE, VAE 2D, VAE 4D and RVAE 10D methods.}
    \vspace{-20pt}
    \label{fig:spreads}
\end{figure}
\section{Results}

The results from the 2D projection calculations with the different methods are shown in Fig.~\ref{fig:spreads}.
For reproducibility, these are generated with a fixed random seed of 42.
The colours of the contours correspond to their given labels.
Additionally, the 4 content word utterances are marked with light yellow halos in the RVAE results, to emphasise its ability to produce a latent space representation of variable length data.
As can be seen in the plots, for this dataset, the t-SNE gives a good spread of the data, which is much better than the results obtained with the PCA, and even the 2D VAE.
The 4D VAE also does a good job at clustering similar contours nearby.
The 10D RVAE is the only method that is able to successfully project longer and shorter contours in a single representations.

In the t-SNE, 4D VAE and 10D RVAE we can also see some misplaced contours. 
This is possibly due to the limitations of projecting high-dimensional data, that can be residing on multiple manifolds in the original high-dimensional space, onto a 2D representation.
Running these methods with different random seeds might give improved results.
This is why the tool allows the user to experiment with multiple representations using the same method and hopefully obtain a good trade-off of between well placed and misplaced contours.
Nevertheless, all of the representations seem to grant an annotator a unique `look' of the data that can aid in the annotation process --- both for labelling unlabelled data and for correcting erroneous labels.


\section{Conclusions}

The proposed ProsoBeast annotation tool is designed to ease the labelling process of prosody rich corpora through the calculation and visualisation of 2D representations of the prosodical latent space of the data.
The tool offers multiple state-of-the-art methods for calculating these projections, each seeking to cluster similar prosodic contours, while distancing dissimilar ones.
The user can use each method a multitude of times to get different representations of their data, and can navigate between them easily via the main screen of the web application.
The usefulness of the tool has been demonstrated with a sample dataset, and it remains to be evaluated on various data by the community.

\section{Acknowledgements}

The design and implementation of the tool was funded by NSERC Grant RGPIN-2018-06153: {\em Three dimensions of sentence prosody} to the second author.

\bibliographystyle{IEEEtran}

\bibliography{mybib}

\begin{thebibliography}{10}
\providecommand{\url}[1]{#1}
\csname url@samestyle\endcsname
\providecommand{\newblock}{\relax}
\providecommand{\bibinfo}[2]{#2}
\providecommand{\BIBentrySTDinterwordspacing}{\spaceskip=0pt\relax}
\providecommand{\BIBentryALTinterwordstretchfactor}{4}
\providecommand{\BIBentryALTinterwordspacing}{\spaceskip=\fontdimen2\font plus
\BIBentryALTinterwordstretchfactor\fontdimen3\font minus
  \fontdimen4\font\relax}
\providecommand{\BIBforeignlanguage}[2]{{%
\expandafter\ifx\csname l@#1\endcsname\relax
\typeout{** WARNING: IEEEtran.bst: No hyphenation pattern has been}%
\typeout{** loaded for the language `#1'. Using the pattern for}%
\typeout{** the default language instead.}%
\else
\language=\csname l@#1\endcsname
\fi
#2}}
\providecommand{\BIBdecl}{\relax}
\BIBdecl

\bibitem{ladd2008intonational}
D.~R. Ladd, \emph{{Intonational Phonology}}, 2nd~ed.\hskip 1em plus 0.5em minus
  0.4em\relax Cambridge University Press, 2008.

\bibitem{pearson1901liii}
K.~Pearson, ``{LIII. On lines and planes of closest fit to systems of points in
  space},'' pp. 559--572, 1901.

\bibitem{hinton2002stochastic}
G.~Hinton and S.~T. Roweis, ``Stochastic neighbor embedding,'' in \emph{NIPS},
  vol.~15.\hskip 1em plus 0.5em minus 0.4em\relax Citeseer, 2002, pp. 833--840.

\bibitem{van2008visualizing}
L.~Van~der Maaten and G.~Hinton, ``Visualizing data using t-{SNE},''
  \emph{Journal of machine learning research}, vol.~9, no.~11, 2008.

\bibitem{cottrell1988principal}
G.~W. Cottrell and P.~Munro, ``Principal components analysis of images via back
  propagation,'' in \emph{Visual Communications and Image Processing'88: Third
  in a Series}, vol. 1001.\hskip 1em plus 0.5em minus 0.4em\relax International
  Society for Optics and Photonics, 1988, pp. 1070--1078.

\bibitem{deng2010binary}
L.~Deng, M.~L. Seltzer, D.~Yu, A.~Acero, A.-r. Mohamed, and G.~Hinton, ``Binary
  coding of speech spectrograms using a deep auto-encoder,'' in \emph{Eleventh
  Annual Conference of the International Speech Communication Association},
  2010.

\bibitem{socher2011semi}
R.~Socher, J.~Pennington, E.~H. Huang, A.~Y. Ng, and C.~D. Manning,
  ``Semi-supervised recursive autoencoders for predicting sentiment
  distributions,'' in \emph{Proceedings of the conference on empirical methods
  in natural language processing}.\hskip 1em plus 0.5em minus 0.4em\relax
  Association for Computational Linguistics, 2011, pp. 151--161.

\bibitem{liou2014autoencoder}
C.-Y. Liou, W.-C. Cheng, J.-W. Liou, and D.-R. Liou, ``Autoencoder for words,''
  \emph{Neurocomputing}, vol. 139, pp. 84--96, 2014.

\bibitem{ap2014autoencoder}
S.~C. Ap, S.~Lauly, H.~Larochelle, M.~Khapra, B.~Ravindran, V.~C. Raykar, and
  A.~Saha, ``An autoencoder approach to learning bilingual word
  representations,'' in \emph{Advances in Neural Information Processing
  Systems}, 2014, pp. 1853--1861.

\bibitem{obin2018sparse}
N.~Obin and J.~Beliao, ``Sparse coding of pitch contours with deep
  auto-encoders,'' in \emph{Speech Prosody}, 2018.

\bibitem{kingma2013auto}
\BIBentryALTinterwordspacing
D.~P. Kingma and M.~Welling, ``Auto-encoding variational {Bayes},''
  \emph{CoRR}, vol. abs/1312.6114, 2013. [Online]. Available:
  \url{http://arxiv.org/abs/1312.6114}
\BIBentrySTDinterwordspacing

\bibitem{fabius2014variational}
O.~Fabius and J.~R. Van~Amersfoort, ``Variational recurrent auto-encoders,''
  \emph{arXiv preprint arXiv:1412.6581}, 2014.

\bibitem{gerazov2018variational}
B.~Gerazov, G.~Bailly, O.~Mohammed, Y.~Xu, and P.~N. Garner, ``A variational
  prosody model for mapping the context-sensitive variation of functional
  prosodic prototypes,'' \emph{arXiv preprint arXiv:1806.08685}, 2018.

\bibitem{scikit-learn}
F.~Pedregosa, G.~Varoquaux, A.~Gramfort, V.~Michel, B.~Thirion, O.~Grisel,
  M.~Blondel, P.~Prettenhofer, R.~Weiss, V.~Dubourg, J.~Vanderplas, A.~Passos,
  D.~Cournapeau, M.~Brucher, M.~Perrot, and E.~Duchesnay, ``Scikit-learn:
  Machine learning in {P}ython,'' \emph{Journal of Machine Learning Research},
  vol.~12, pp. 2825--2830, 2011.

\bibitem{wagnergoodhue21}
\BIBentryALTinterwordspacing
M.~Wagner and D.~Goodhue, ``Toward a bestiary of {English} intonational tunes:
  {Data},'' 2021, {OSF} Project. [Online]. Available:
  \url{https://doi.org/10.17605/OSF.IO/H8DYA}
\BIBentrySTDinterwordspacing

\bibitem{goodhue2016toward}
D.~Goodhue, L.~Harrison, Y.~C. Su, and M.~Wagner, ``Toward a bestiary of
  {English} intonational contours,'' \emph{The Proceedings of the North East
  Linguistics Society (NELS)}, vol.~46, pp. 311--320, 2016.

\bibitem{goodh18}
D.~Goodhue, ``Polarity focus as focus,'' \emph{Journal of Semantics},
  submitted.

\bibitem{liber74}
M.~Liberman and I.~Sag, ``Prosodic form and discourse function,'' in
  \emph{Proceedings of {CLS}}, vol.~10, 1974, pp. 416--427.

\bibitem{ward85}
G.~Ward and J.~Hirschberg, ``Implicating uncertainty: The pragmatics of
  fall-rise intonation,'' \emph{Language}, vol.~61, no.~3, pp. 747--776, 1985.

\bibitem{hirsc92}
J.~Hirschberg and G.~L. Ward, ``The influence of pitch range, duration,
  amplitude and spectral features on the interpretation of the rise-fall-rise
  intonation contour in {English},'' \emph{Journal of Phonetics}, vol.~20, pp.
  241--251, 1992.

\bibitem{ghahremani2014pitch}
P.~Ghahremani, B.~BabaAli, D.~Povey, K.~Riedhammer, J.~Trmal, and S.~Khudanpur,
  ``A pitch extraction algorithm tuned for automatic speech recognition,'' in
  \emph{Proceedings of the IEEE International Conference on Acoustics, Speech
  and Signal Processing}.\hskip 1em plus 0.5em minus 0.4em\relax IEEE, 2014,
  pp. 2494--2498.

\bibitem{hirst2011analysis}
D.~Hirst, ``The analysis by synthesis of speech melody: from data to models,''
  \emph{Journal of speech Sciences}, vol.~1, no.~1, pp. 55--83, 2011.

\bibitem{boersma2001praat}
P.~Boersma, ``Praat, a system for doing phonetics by computer,'' \emph{Glot.
  Int.}, vol.~5, no.~9, pp. 341--345, 2001.

\bibitem{rosenblatt1956remarks}
M.~Rosenblatt, ``Remarks on some nonparametric estimates of a density
  function,'' \emph{The Annals of Mathematical Statistics}, vol.~27, no.~3, pp.
  832 -- 837, 1956.

\bibitem{kingma2014adam}
\BIBentryALTinterwordspacing
D.~P. Kingma and J.~Ba, ``Adam: {A} method for stochastic optimization,''
  \emph{CoRR}, vol. abs/1412.6980, 2014. [Online]. Available:
  \url{http://arxiv.org/abs/1412.6980}
\BIBentrySTDinterwordspacing

\end{thebibliography}

\end{document}